\documentclass[aps,pre,reprint,showpacs,groupedaddress,amsmath,amssymb]{revtex4-1}

\usepackage{graphicx}

\usepackage{color}

\makeatletter
\let\start@align@nopar\start@align
\let\start@gather@nopar\start@gather
\let\start@multline@nopar\start@multline
\long\def\start@align{\par\start@align@nopar}
\long\def\start@gather{\par\start@gather@nopar}
\long\def\start@multline{\par\start@multline@nopar}
\makeatother
\begin{document}

%\preprint{}

\title{Distribution of population averaged observables in stochastic gene expression}


\author{Bhaswati Bhattacharyya}
\email[]{bbhattacharyya@icems.kyoto-u.ac.jp}
%\thanks{}

\author{Ziya Kalay}
\email[]{zkalay@icems.kyoto-u.ac.jp}

\affiliation{Institute for Integrated Cell-Material Sciences (WPI-iCeMS), Kyoto University, Yoshida Ushinomiya-cho, Sakyo-ku, 606-8501, Japan}


\date{\today}

\begin{abstract}
Observation of phenotypic diversity in a population of genetically identical cells is often linked to the stochastic nature of chemical reactions involved in gene regulatory networks. We investigate the distribution of population averaged gene expression levels as a function of population, or sample, size for several stochastic gene expression models to find out to what extent population averaged quantities reflect the underlying mechanism of gene expression. We consider three basic gene regulation networks corresponding to transcription with and without gene state switching and translation. Using analytical expressions for the probability generating function of observables and Large Deviation Theory, we calculate the distribution and first two moments of the population averaged mRNA and protein levels as a function of model parameters, population size and number of measurements contained in a data set. We validate our results using stochastic simulations also report exact results on the asymptotic properties of population averages which show qualitative differences among different models.
\end{abstract}

\pacs{82.39Rt, 82.39.-k, 82.20.Db, 02.50.-r}
%\keywords{}

\maketitle

\section{Introduction}
\label{sec:introduction}
One of the recent puzzles in cell biology is quantifying and interpreting the phenotypic heterogeneity in a population of genetically identical cells~\cite{eldar_functional_2010,huang_non-genetic_2009,lionnet_transcription_2012}. With the help of advanced biochemistry and microscopy techniques, it has been possible to measure gene expression at the single cell level~\cite{larson_single_2009}. This led to the finding that genetically identical cells can greatly differ in their gene expression profiles, a phenomenon sometimes referred as \emph{population heterogeneity}~\cite{huang_non-genetic_2009}, and that a single cell's gene expression pattern can significantly change over time~\cite{elowitz_stochastic_2002, hirata_oscillatory_2002,spiller_measurement_2010}. While this apparently heterogeneous and dynamic nature of cells had often been overlooked as a source of \emph{noise} in measurements, it is now becoming clear that the stochastic nature of gene expression plays a fundamental role in processes such as stem cell differentiation~\cite{singh_heterogeneous_2007, kalmar_regulated_2009, imayoshi_oscillatory_2013}, improving the overall fitness of single-celled organisms~\cite{sharma_chromatin-mediated_2010}, and more~\cite{raj_nature_2008, geiler-samerotte_details_2013}.

The ground-breaking experimental findings of the last decade, spearheaded by the observation of transcription at the single mRNA level~\cite{raj_stochastic_2006}, has stimulated a large number of theorists to develop quantitative models to gain insight into the stochastic nature of gene expression~\cite{paulsson_summing_2004,walczak_analytic_2012}. The majority of these works consist of modeling the synthesis of mRNA and the subsequent protein as stochastic processes, such that the predictions consist of probability distributions for the number of mRNA and protein molecules at a single cell level. For a recent review on how gene expression noise can be used to infer regulation mechanisms see, for instance, Munsky et al.~\cite{munsky_using_2012}. Theoretical predictions have often been compared against experimental data in bacteria~\cite{golding_real-time_2005,taniguchi_quantifying_2010}, and new techniques are being developed to explore gene expression patterns in mammalian cells at the level of single molecules, which is much more challenging.

\begin{figure}[h]
\centering
\includegraphics[width=0.9\columnwidth]{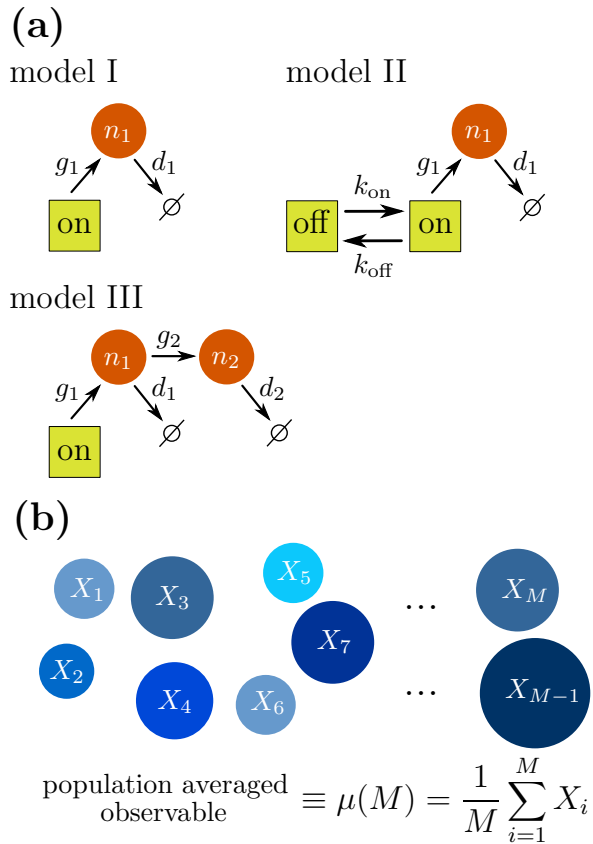}
\caption{(Color online) Schematic illustration of the three models of gene expression and of population averaging. (a) In all three biochemical networks squares (yellow) denote the gene, which can be in the on or off state. The circles (brown) represent mRNA and protein molecules, whose numbers are equal to $n_1$  and $n_2$, respectively, and the arrows into $\emptyset$ correspond to degradation. In model I, the gene is always on and mRNA is generated and degraded with the rates $g_1$ and $d_1 n_1$, respectively. model II describes a gene which is switching to the on (off) state with the rate $k_{\mathrm{on}}$ ($k_{\mathrm{off}}$). While the gene is on, mRNA is produced at the rate $g_1$. In model III, the gene is always on and mRNA is generated with the rate $g_1$, which subsequently produces proteins with the rate $g_2 n_1$. (b) The expression level of each cell is denoted as a random variable $X_i$ such that the population averaged observable is the sample mean over $M$ cells. For models I and II, $X_i=n_1$ and for model III, $X_i=n_2$. $X_i$'s are assumed to be distributed identically and independently.}
\label{fig:figure1}
\end{figure}

In this work, we study the distribution of \emph{population averaged} gene expression levels for several basic but fundamental gene expression models. In our definition, a population averaged measurement corresponds to that obtained by averaging over the signal collected from a subset of a colony of cells, where the signal is commonly the fluorescence from reporter molecules whose numbers are proportional to the mRNA/protein levels (see Fig. \ref{fig:figure1} for an illustration). As we mentioned above, the majority of previous theoretical work in this subject centered on calculating the distribution of mRNA and protein molecules in single cells. Nevertheless, many experimental techniques that can quantify gene expression are performed using a large number of cells, providing population averaged observables. For instance, in commonly used techniques employing microarray analysis~\cite{hoheisel_microarray_2006} and real-time polymerase chain reaction (or qPCR)~\cite{higuchi_pcr_1993} expression at the population level constitutes the readout, and information at the single cell level is nearly lost. However, these techniques usually require a sample that consists of millions of cells, and it is costly to repeat the experiments to construct probability distributions for the outcome. Hence, it is usually not possible to go beyond the measurement of mean expression level that is not very informative about the mechanism of gene expression. On the other hand, in cases where the quantity of mRNA or proteins inside cells can be monitored by a fluorescent reporter, flow cytometry~\cite{givan_flow_2001} can be used as a powerful tool to monitor a large number of cells. Flow cytometry makes it possible to record the fluorescent signal from each cell in a population, or record sample averaged fluorescent signal by averaging over a controlled number of cells. To analyze gene expression patterns for sets of millions of cells, it might be beneficial to record sample averages to avoid the accumulation of huge data sets. In such an approach, the key question that arises is: given the precision of experimental measurement, how large the sample can get while the measurements still provide information beyond the average expression level?

Here we consider three well-known models for stochastic gene expression describing certain aspects of transcription and translation to make predictions about the distribution and moments of population averaged mRNA and protein levels, and to assess the feasibility of using population averages to infer the properties of a single cell by addressing the question we posed above.

The article is organized as follows. In section \ref{sec:background}, we present the details of the stochastic gene expression models considered in this work (see Fig. \ref{fig:figure1}(a)) along with their mathematical formulation. In section \ref{sec:population_averages}, we describe how the population averages are calculated from probability generating functions via exact methods and the Large Deviation Theory, and we present our results in section \ref{sec:specific_results}. Lastly, we discuss the findings and present our conclusions in section \ref{sec:discussion}.

\section{Gene expression models and their mathematical formulation}
\label{sec:background}
We consider three different biochemical networks that are often used to model basic processes during gene expression. A schematic illustration of these models is given in Fig. \ref{fig:figure1}(a). Models I and II are concerned with the transcription process for constitutively and transiently expressed genes, respectively, and model III captures the transcription and translation processes for a constitutively expressed gene. We are interested in studying the predictions of these models in regimes where the concentration of molecules is small such that a stochastic description of chemical kinetics is required. In this respect, we consider the Chemical Master Equation (CME) for these three models, and work with its analytical solutions as we explain below. In the rest of this section, we provide the mathematical formulation of these three models and present the results that are needed to calculate the quantities of our interest.

Model I has been used to calculate the distribution of mRNA molecules when a gene is expressed constitutively, that is when the gene is always turned on over the time scale of observations. In bacteria and yeast cells where the typical numbers of mRNA molecules at steady state could be close to zero, predictions of this model can be comparable with experimental findings~\cite{taniguchi_quantifying_2010}. The CME for this model is given by
\begin{align}
\frac{dP_{\mathrm{I}}(n_1,t)}{dt} &= g_1P_{\mathrm{I}}(n_1-1,t) - (g_1+d_1 n_1)P_{\mathrm{I}}(n_1,t) \nonumber \\ &+ d_1(n_1+1)P_{\mathrm{I}}(n_1+1,t),
\label{eqn:CME_model_I}
\end{align}
where $P_{\mathrm{I}}(n_1,t)$ is the probability to find $n_1$ molecules of mRNA at time $t$, and in general we will use $P_{i}(\vec{n},t)$ to denote the analogous probability for model $i$. It is well known that the solution of Eq. (\ref{eqn:CME_model_I}) at steady state is a Poisson distribution (also see Appendix \ref{sec:appendix_2}). In our approach, we make use of the probability generating function (pgf) for a model to be able to calculate the distribution of population averaged observables, where the pgf is formally defined as
\begin{gather}
Q(\vec{z}) = \sum_{\vec{n}=0}^{\infty} \prod_{i}z_i^{n_i} P(\vec{n}).
\label{eqn:pgf}
\end{gather}
For model I, the pgf is given by
\begin{gather}
Q_{\mathrm{I}}(z) = e^{-\mu_{\mathrm{I}}^{0}(1-z)},
\label{eqn:poisson_pgf}
\end{gather}
where $\mu_{\mathrm{I}}^{0}=g_{1,1}$ and we use the short hand notation $g_{i,j}=g_i/d_j$ and $d_{i,j}=d_i/d_j$ throughout the article. The mean and variance of $n_1$ are given by
\begin{gather}
\mu_{\mathrm{I}}^0 = g_{1,1}, \hspace{0.5cm} {\sigma^0_{\mathrm{I}}}^2 = \mu_{\mathrm{I}}^0.
\end{gather}

Model II is an extension of model I to the case where a gene can transition between on and off states, where no mRNA molecule is synthesized in the off state. This model is also known as the \emph{random telegraph model}, introduced by Ko~\cite{ko_stochastic_1991} based on early experimental observations. If the rate of transition from the on state to the off state is equal to $k_{\textrm{off}}$, and $k_{\textrm{on}}$ vice versa, the CME for this model can be written as
\begin{align}
&\frac{dP_{\mathrm{II}}(n_1,s,t)}{dt} = g_1 s P_{\mathrm{II}}(n_1-1,s,t) \nonumber \\ &+ d_1(n_1+1)P_{\mathrm{II}}(n_1+1,s,t) \nonumber \\  &+ k_{\rm off} (1-s) P_{\mathrm{II}}(n_1,s,t) \nonumber \\ &+ k_{\rm on} s P_{\mathrm{II}}(n_1,1-s,t)   \nonumber \\ &- (g_1 + d_1 n_1 + k_{\rm off} s + k_{\rm on} (1-s) )P_{\mathrm{II}}(n_1,s,t),
\label{eqn:CME_model_II}
\end{align}
where $s=1$ when the gene is in the on state and 0 when it is off. The pgf for this model at steady state was given by Peccoud and Ycart~\cite{peccoud_markovian_1995} as
\begin{align}
Q_{\mathrm{II}}(z) &= \sum_{n_1=0}^{\infty}\sum_{s = 0}^{1}z^{n_1} P_{\mathrm{II}}(n_1,s) \nonumber \\ &= {}_1F_1\left(\frac{k_{\textrm{on}}}{d_1}, \frac{k_{\textrm{on}}+k_{\textrm{off}}}{d_1}, g_{1,1}(z-1) \right),
\label{eqn:pec_yca_pgf}
\end{align}
where ${}_1F_1(a,b,z)$ is the Kummer confluent hypergeometric function~\cite{abramowitz_handbook_1964}. For a generalization of this result that allows for non-Markovian transitions between the on and off states, see Stinchcombe et al.~\cite{stinchcombe_population_2012}. For model II, the mean and variance of $n_1$ is given by
\begin{align}
\mu_{\mathrm{II}}^0 &=\frac{g_{1,1} k_{\rm on}}{(k_{\rm on} + k_{\rm off})}, \\
{\sigma^0_{\mathrm{II}}}^2 &= \mu_{\mathrm{II}}^0 \left[ 1 + \frac{g_{1,1} k_{\mathrm{off}} \left( k_{\mathrm{on}} + k_{\mathrm{off}} \right)^{-1}} {1 + \left( k_{\mathrm{on}} + k_{\mathrm{off}} \right) d_1^{-1}}\right]. 
\end{align}

Model III describes the synthesis of a molecule in two steps, and has been used to study protein expression. It captures the \emph{burstiness} of protein production, as a single mRNA molecule can be translated several times during its lifetime, resulting in a bunch of proteins. This model has also been extensively compared with experimental data in studies of protein expression in bacteria and yeast. In this case the CME is given by
\begin{align}
&\frac{dP_{\mathrm{III}}(n_1,n_2,t)}{dt} = g_1P_{\mathrm{III}}(n_1-1,n_2,t) \nonumber \\ &+ d_1(n_1+1)P_{\mathrm{III}}(n_1+1,n_2,t) \nonumber \\  &+ g_2 n_1 P_{\mathrm{III}}(n_1,n_2-1,t) \nonumber \\ &+ d_2 (n_2+1)P_{\mathrm{III}}(n_1,n_2+1,t)   \nonumber \\ &- (g_1 + d_1 n_1 + g_2 n_1 +d_2 n_2 )P_{\mathrm{III}}(n_1,n_2,t),
\label{eqn:CME_model_III}
\end{align}
and the corresponding pgf for the protein distribution at steady state is found to be (see Appendix \ref{sec:appendix_2} for alternative forms)
\begin{gather}
Q_{\mathrm{III}}(z) = \exp \Bigg[ g_{1,2} \int_0^{g_{2,2}(z-1)} dv e^{v} \frac{\gamma(d_{1,2},v)}{v^{d_{1,2}}} \Bigg],
\label{eqn:pgf_model_3_intro}
\end{gather}
where $\gamma(a,z)$ is the lower incomplete gamma function defined as
\begin{gather*}
\gamma(a,z)=\int_0^z dt e^{-t} t^{a-1}.
\end{gather*}
The derivation of Eq. (\ref{eqn:pgf_model_3_intro}), along with the joint probability generating function $Q_{\mathrm{III}}(z_1,z_2)$, is outlined in Appendix \ref{sec:appendix_2}, based on the work by Shahrezaei and Swain~\cite{shahrezaei_analytical_2008} but without making the approximation that the mRNA lifetime is negligible compared with that of the protein (also see Bokes et al.~\cite{bokes_exact_2012} and Pendar et al.~\cite{pendar_exact_2013} for alternative derivations). The mean and variance of the final product $n_2$ are equal to
\begin{gather}
\mu_{\mathrm{III}}^0 =g_{1,1}g_{2,2}, \hspace{0.5cm} {\sigma^0_{\mathrm{III}}}^2 = \mu_{\mathrm{III}}^0\left(1 + \frac{g_{2,1}}{1+d_{2,1}} \right).
\end{gather}

\section{Calculation of population averages}
\label{sec:population_averages}
In this section we present how population averaged observables are calculated. We consider the case where each cell in the population can be thought as an independent chemical system, which could be a reasonable approximation if cell-to-cell communication negligibly alters the expression of the monitored gene product. 

We envisage a measurement in which the observable is the average amount of mRNA or protein expressed by $M$ cells, as illustrated in Fig. \ref{fig:figure1}(b). At the time of measurement, we suppose that the amount of mRNA or protein in the $i^{\rm th}$ cell is a random variable $X_i$. Let $\mu(M)$ denote the population averaged output from these $M$ cells, defined by
\begin{gather}
\mu(M) = \frac{1}{M}\sum_{i=1}^{M} X_{i}.
\label{eqn:sample_mean}
\end{gather}
If $N\gg 1$ independent measurements of $\mu(M)$ are performed, one can construct the probability distribution for $\mu(M)$, which contains information about higher moments of $X$.

\subsection{Exact calculations}
In some simple cases, the distribution of $\mu(M)$ can be obtained exactly. In this subsection, we outline the method of obtaining the exact distributions and moments of $\mu(M)$.
    
Since $X_i$'s are independently and identically distributed, their sum, defined as a random variable $Y(M)=\sum_{i=1}^M X_i$, obeys a distribution whose generating function is the $M^{\mathrm{th}}$ power of the probability generating function for $X_i$. Therefore, we have 
\begin{gather}
Q_Y\left( z; M\right) = Q_X\left( z\right)^M.
\label{eqn:Q_Y_Q_X}
\end{gather}
Note that $Y(M)$ is an integer valued random variable, contrary to $\mu(M)$ which can take any positive real value. We can obtain the probability distribution function of $\mu(M)$ by calculating the probability that $Y(M)$ lies in intervals bounded by integers. For convenience, we choose a binsize of $M$, such that the $m^{\mathrm{th}}$ bin, centered at $Mm$, corresponds to $Y(M)$ being between $L_m=\lceil \max((m-0.5)M,0) \rceil$ and $R_m = \lceil (m+0.5)M-1 \rceil$, such that $ \lceil \max((m-0.5),0)  \rceil \leq \mu(M) <  \lceil m+0.5 \rceil $, where $m=0,1,2,...$ . Therefore, the probability distribution of $\mu(M)$ is given by
\begin{gather}
P_{\mu}(m;M) = \sum_{y= L_m }^{ R_m } P_Y( y),
\label{eqn:P_mu_formal}
\end{gather}
which we use as a shorthand notation for $P_{\mu}( \lceil \max((m-0.5),0)  \rceil \leq \mu <  \lceil m+0.5 \rceil   ;M)$.

Due to the algebraic complexity of the calculations, it might not always be plausible to calculate the full distribution given in Eq. (\ref{eqn:P_mu_formal}). More commonly, one can calculate the first few moments of $\mu(M)$ exactly. Knowing the probability generating function, the $\ell^{\mathrm{th}}$ moment of $\mu(M)$ can be calculated as 
\begin{gather}
\langle \mu^{\ell} \rangle = \frac{1}{M^{\ell}} \lim_{k \to 0} \frac{d^{\ell}}{d k^{\ell}} Q_Y(e^{k y};M),
\end{gather}
when the differentiation and limit can be performed conveniently. Otherwise, it may be preferable to express the moments of $Y(M)$ in terms of the moments of $X$, which are easier to calculate as long as the random variables $X_{i}$ are independently and identically distributed~\cite{packwood_moments_2011} . 

Due to measurement uncertainties and the limited number of data points, it is usually challenging to determine higher moments of an observable obtained via experiments. As a result, one can usually calculate only the first few moments accurately enough to do further analysis. As the intrinsic variance of observables contain information about the underlying stochastic process, it is important to determine whether the observed variance is due to measurement uncertainty or not. Measurement uncertainty can be separated into two parts: 1) uncertainty due to the finite number of data points, and 2) uncertainty due to systematic measurement errors. Usually the systematic error is not straightforward to assess. In this work, we would like to discuss the effect of having a finite number of data points on the variance of calculated moments, which is relevant for all experiments. We restrict our analysis to the variance of the mean and variance of the observable; however, one can extend the calculations for arbitrarily higher moments at the expense of lengthy algebra.

Unbiased estimators for the mean and variance of a sample obtained by $N$ independent measurements are given by 
\begin{align}
m(N) &= \frac{1}{N}\sum_{i=1}^{N} {\mu(M)}_i, \label{eqn:estimator_mu} \\
{s(N)}^2 &= \frac{1}{N-1} \sum_{i=1}^N \left( {\mu(M)}_i - m(N) \right)^2, \label{eqn:estimator_sigma}
\end{align}
where the expected values of $m(N)$ and $s(N)$ (i.e. as $N\to \infty$) are equal to $\mu^{0}(M)$ and $\sigma^{0}(M)$, respectively, which are the mean and standard deviation of the random variable $\mu(M)$. Using Eq. (\ref{eqn:estimator_mu}) and (\ref{eqn:estimator_sigma}), we can express the uncertainty in the estimation of the mean as a function $M$ and the number of data points as
\begin{gather}
\mu(M,N) = \mu^0 \pm \frac{\sigma^0}{\sqrt{MN}}, 
\label{eqn:mu_estimate}
\end{gather}        
where the uncertainty is equal to the standard deviation of $\mu (M,N)$ around its mean value, $\mu^0=\mu^{0}(1)$, and $\sigma^0=\sigma^{0} (1)$. To estimate the finite size effects in the measurement of the variance, we need to calculate the expectation value
\begin{gather}
\left \langle \left( {s(N)}^2 - {\sigma^{0}(M)}^2 \right)^2 \right \rangle,
\end{gather}
which corresponds to a variance of variance. Going through straightforward but lengthy algebra, we obtain the above term, and express the effect of a finite data points on the value of the variance as
\begin{align}
{\sigma (M,N)}^2 &= \frac{{\sigma^{0}}^2}{M} \nonumber \\ &\pm \sqrt{ \frac{ \left \langle \left( \mu(M) - \mu^{0} \right)^4 \right \rangle}{N} - \frac{{\sigma^{0}}^4 (N-3)}{M^2 N(N-1)} },
\label{eqn:var_estimate}
\end{align}
which involves the fourth centralized moment of $\mu(M)$.
 
When a detailed knowledge about the statistics of the systematic uncertainty is available, one can put tighter bounds on the number of data points needed to achieve a desired level of accuracy.

\subsection{Approximation via the Large Deviation Theory}

For each of the three models, we calculate the predicted distribution of $\mu_i(M)$, where $i$ is the model index. In order to calculate the distribution of population averaged observables, which are sample means, we make use of several results from the Large Deviation Theory (LDT). The LDT provides a convenient framework for calculating the distribution of a random variable which becomes more and more concentrated around its mean value as a certain parameter in the system is taken to infinity (i.e., when the central limit theorem holds). For instance, in equilibrium statistical physics, this ``certain parameter'' could be number of molecules in an ideal gas under constant pressure and temperature so that the limit corresponds to the thermodynamic limit, where the energy of the system can be treated as a deterministic value even though the energy of its constituents are constantly fluctuating. For our purpose, we are interested in using LDT to calculate the distribution of $\mu_i(M)$, which concentrates around the value $\mu_i^0$ in the limit of large number of cells, i.e. $M\to \infty$. 

Without going into details, here we quote a key theorem in LDT with which we calculate the distribution of the population averaged observable $\mu (M)$. For an excellent review of the subject and technical details, we refer the reader to refs.~\cite{ellis_theory_1999, touchette_large_2009}. According to the LDT, the distribution of a random variable $A_n$, whose values become increasingly concentrated around a certain value as $n\to \infty$, can be approximated by
\begin{gather}
P\left( A_n \in \left[ a, a+da \right] \right) \simeq e^{-n I(a)}da,
\label{eqn:probability_distribution}
\end{gather}
where $I(a)$ is often called the rate function. The rate function can be calculated via the G{\"a}rtner-Ellis theorem~\cite{ellis_theory_1999}, which states that $I(a)$ is the Legendre-Fenchel transform of the scaled cumulant generating function of $A_n$, given by
\begin{gather}
\lambda(k) = \lim_{n\to \infty} \frac{1}{n} \ln \left\langle e^{n k A_n} \right\rangle,
\end{gather}
and the Legendre-Fenchel transform is defined as
\begin{gather}
I(a) = \sup_{k \in \mathbb{R}} \left\{ ka - \lambda(k) \right\},
\end{gather}
where sup stands for limit supremum. In our case, the variable $A_n$ is equivalent to the sample mean $\mu(M)$ defined in Eq. (\ref{eqn:sample_mean}), where $M$ plays the role of the parameter $n$. For sample means, the function $\lambda(k)$ is given by~\cite{touchette_large_2009}
\begin{gather}
\lambda(k) = \ln Q(e^k),
\label{eqn:equation_for_lambda}
\end{gather}
where $Q(z)$ is the pgf of the identically and independently distributed random variables $X_i$ in the summation on the right hand side of Eq. (\ref{eqn:sample_mean}). When the sample mean converges to a unique value $\mu^0$ as $M\to \infty$, calculating $I(\mu)$ reduces to
\begin{gather}
I(\mu) = k^* \mu - \lambda(k^*),
\label{eqn:equation_for_I}
\end{gather}
where $k^*$ is the root of
\begin{gather}
\frac{d \lambda}{d k} = \frac{d \ln[Q(e^k)]}{d k} =\mu,
\label{eqn:for_the_root}
\end{gather}
%Which can be rewritten as 
%\begin{gather}
%\frac{d \ln[Q(e^k)]}{d k} = \mu.
%\label{eqn:for_the_root_2}
%\end{gather}
where $Q$ for models I, II and III are given by Eqs. (\ref{eqn:poisson_pgf}, \ref{eqn:pec_yca_pgf}) and (\ref{eqn:pgf_model_3_intro}).

For the basic biochemical networks considered in this article, we can obtain some of the properties of the function $I(\mu)$, and hence the probability distribution of $\mu(M)$, via analytical or numerically exact calculations. In the following section, we display the outcomes for specific cases.

%%%%%%%%%%%%%%%%%%%%%%RESULTS%%%%%%%%%%%%%%%%%%%%%%%%%%%
\section{Results}
\label{sec:specific_results}

\subsection{Estimating the first two moments and the Fano factor}

When it is appropriate to fit a model to data, it is often possible to compare the first few moments calculated from the data to the model predictions. This is because one needs larger and larger data sets to accurately calculate higher moments, which is usually not available. In this section we present our results on how the uncertainty in the mean and variance of $\mu(M)$ depends on $M$ and the number of independent measurements $N$, and discuss the estimation of the Fano factor derived from these statistics.   

In order to estimate $\mu(M)$ with accuracy, uncertainty due to sampling should be small, such that $\mu^0 \gg (MN)^{-1/2}\sigma^0$ (see Eq. (\ref{eqn:mu_estimate})). Therefore, large values of both $M$ and $N$ are favorable, since the uncertainty decreases either way. In contrast, to estimate ${\sigma^0}^2$ in the presence of experimental errors, $M$ cannot be arbitrarily large. As $\sigma(M,N)^2$ decays linearly with $M$, there will be a certain value of $M$ where the precision of measurements will be comparable to the expected value of $\sigma(M,N)^2$, that is ${\sigma^0}^2/M$ (see Eq. (\ref{eqn:var_estimate})). At this point, population averaged measurements will cease to be informative as we cannot calculate second and higher moments accurately.

A commonly used quantity for analyzing the burstiness of gene expression is the Fano factor defined as
\begin{gather}
\phi = {\sigma^2}/{\mu}.
\label{eqn:fano_factor_formal}
\end{gather}
Among the models we consider, model I corresponds to a process where the output is not bursty such that $\phi=1$, whereas the other models produce bursty output resulting in $\phi \geq 1$. In line with our discussion above, to estimate $\phi$ accurately, $M$ has to satisfy
\begin{gather}
\frac{{\sigma^0}^2}{\mu^0 M} \gg \Delta \mathcal{I},
\end{gather}
where $\Delta \mathcal{I}$ is the precision at which signal, for instance fluorescence intensity from reporter molecules, can be measured. Therefore, the measurement precision puts an upper bound on the value of $M$. Noting that the value of $\phi$ can be close to 1, and that the values of ${\sigma^0}^2$ and $\mu^0$ are not available a priori, a conservative estimate of the maximum value of $M$ can be stated as
\begin{gather}
M_{\mathrm{max}} \lesssim \Delta \mathcal{I} ^{-1}.
\label{eqn:M_max}
\end{gather}
Using Eqs. (\ref{eqn:mu_estimate}, \ref{eqn:var_estimate}) and (\ref{eqn:fano_factor_formal}), we can express the ratio of the uncertainty in measuring $\phi$, denoted by $\Delta \phi$, to its expected value as
\begin{gather}
\frac{\Delta \phi}{\phi} = \sqrt{\frac{2}{N-1} + \frac{f}{MN}},
\label{eqn:fano_uncertainty}
\end{gather}
where we used the well-known relation ${\Delta z}/{z} = \left( ({\Delta x}/{x})^2 + ({\Delta y}/{y})^2  \right)^{1/2}$ for the propagation of uncertainty in the division $z=x/y$. In Eq. (\ref{eqn:fano_uncertainty}), the term $f$ is a function of all the model parameters and does not depend on either $M$ or $N$. When $f/M \ll 1$, the uncertainty in $\phi$ is dominated by the first term under the square root sign. In such a case, a measurement containing, for instance, $N=100$ data points would result in a relative uncertainty of $\sim 0.14$.

\begin{figure}[h]
\includegraphics[width=8cm]{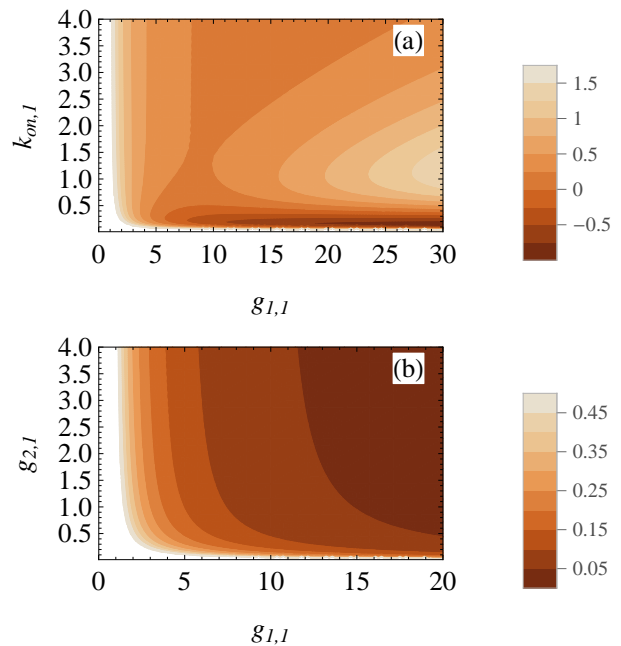}
\caption{(Color online) Contour plots of the term $f$ appearing in Eq. (\ref{eqn:fano_uncertainty}) for models II and III. Results for: (a) model II, for a fixed value of $k_{\mathrm{off},1}=0.1$, and (b) for model III, with $d_{2,1}=0.1$. All rates are measured in units of $d_1$.}
\label{fig:var_var}
\end{figure}

To demonstrate how $f$ depends on model parameters, in Fig. \ref{fig:var_var} we plotted $f$ as a function of two independently varied parameter combinations for models II and III. In order to calculate $f$ via Eqs. (\ref{eqn:mu_estimate}) and (\ref{eqn:var_estimate}) one needs to evaluate the moments of $\mu(M)$ up to the fourth order. In Appendix~\ref{sec:appendix_0} we provide relevant expressions in some detail.

In Fig. \ref{fig:var_var}(a), the contour plot of $f$ for model II is plotted while keeping $k_{\mathrm{off},1}$ constant at $0.1$, and varying $g_{1,1}$ and $k_{\mathrm{on},1}$ (we use the convention $k_{i,j}=k_{i}/d_{j}$). We observe that $f$ behaves non-monotonically as a function of the rate of transcription and the rate at which the gene transitions into the on state. For $g_{1,1}\sim 0$, the behavior of $f$ quickly saturates with increasing $k_{\mathrm{on},1}$, as $\phi$ asymptotes to its minimum value of 1. In this limit, models I and II would be comparable. For this parameter set, when $k_{\mathrm{on},1} \ll 1$ and $k_{\mathrm{on},1} \gg 1$, $f$ decreases with increasing $g_{1,1}$ as the mean expression level increases faster than its variance, resulting in smaller values of the relative uncertainty in $\phi$. Nevertheless, in the intermediate regime $f$ behaves in the opposite manner, increasing with $g_{1,1}$. We observed that the extent of this intermediate regime increases with the value of $k_{\mathrm{off},1}$ indicating that increased variability in expression levels lead to increased uncertainty in measuring $\phi$ as one would expect. In Fig. \ref{fig:var_var}(b), we display a similar plot for model III with $d_{2,1}=0.1$, corresponding to a case where the degradation rate of the protein is an order of magnitude lower than that of the mRNA. In this case $f$ decreases monotonically with the variables $g_{1,1}$ and $g_{2,1}$, which could be explained by the decreased variability of the expression level with respect to its mean, in the presence of a large number of molecules.

\subsection{Results obtained via LDT}
In this section, we present our results comparing $P_{\mu}(m;M)$ calculated exactly or numerically with that obtained by using LDT. We begin by discussing results regarding model I, for which it is possible to calculate $P_{\mu}$ via LDT up to quadratures. After that, we move on to discuss results for the other models and asymptotic properties of the corresponding rate functions $I_i(\mu)$.

\subsubsection{Exact results for model I}

In the case of model I, we can calculate the rate function $I_{\mathrm{I}}(\mu)$ exactly, as follows. Substituting the pgf given in Eq. (\ref{eqn:poisson_pgf}) into Eq. (\ref{eqn:equation_for_lambda}), solving for $k^*$ via Eq. (\ref{eqn:for_the_root}) and using the result in Eq. (\ref{eqn:equation_for_I}), we obtain
\begin{gather}
I_{\mathrm{I}}(\mu) = \mu\ln\left(\mu/\mu^0_{\mathrm{I}}\right) - \mu + \mu_{\mathrm{I}}^0.
\end{gather}
Using Eq. (\ref{eqn:probability_distribution}), we arrive at the LDT approximation of the probability density
\begin{gather}
P(\mu \in [\mu , \mu + d\mu]) \simeq e^{-MI_{\mathrm{I}}(\mu)} d\mu.
\label{eqn:model1_pdf}
\end{gather}
Due to the simplicity of the pgf, we can also calculate the distribution of $\mu_{\mathrm{I}}(M)$ exactly, through Eqs. (\ref{eqn:Q_Y_Q_X}) and (\ref{eqn:P_mu_formal}). Substituting Eq. (\ref{eqn:poisson_pgf}) in Eq. (\ref{eqn:Q_Y_Q_X}), inverting the generating function and using the result in Eq. (\ref{eqn:P_mu_formal}), we obtain the exact distribution as
\begin{gather}
P_{\mu}^{\mathrm{I, exc}}(m;M) =  e^{-M\mu_{\mathrm{I}}^0}\sum_{n=L_m}^{R_m} \frac{(M\mu_{\mathrm{I}}^0)^n}{n!},
\label{eqn:model1_exact}
\end{gather}
where $L_m=\lceil \max((m-0.5)M,0) \rceil$ and $R_m = \lceil (m+0.5)M-1 \rceil$

In order to compare Eq. (\ref{eqn:model1_exact}) with the results obtained by LDT, we integrate Eq. (\ref{eqn:model1_pdf}) over each interval bounded by $L_m$ and $R_m$, and normalize, to obtain
\begin{gather}
P_{\mu}^{\mathrm{I, LDT}}(m;M) = A^{-1} \int_{L_m}^{R_m} d\mu e^{-MI_{\mathrm{I}}(\mu)},
\end{gather}
where $A=\int_0^{\infty} d\mu e^{-MI_{\mathrm{I}}(\mu)}$.

In order to quantify the difference between the exact distribution and the LDT prediction, we employ the Jensen-Shannon divergence~\cite{lin_divergence_1991}, defined as
\begin{align}
\mathrm{JSd}(P_1,P_2) &= \frac{1}{2}\Bigg[ F\left(P_1, \frac{P_1+P_2}{2}\right) \nonumber \\ &+ F\left(P_2, \frac{P_1+P_2}{2} \right) \Bigg], \label{eqn:JSd} \\
F(P_1,P_2) &= \sum_i P_1(i) \ln \frac{P_1(i)}{P_2(i)}. \nonumber
\end{align}
JSd is the average of the relative entropy between two distributions and their average, and can be used as a measure of how close two distributions are. If the two distributions $P_1$ and $P_2$ are identical, $\mathrm{JSd}=0$, and its value grows as the distributions start to differ, with an upper bound of $\ln 2$. 

\begin{figure}[h]
\centering
\includegraphics[width=\columnwidth]{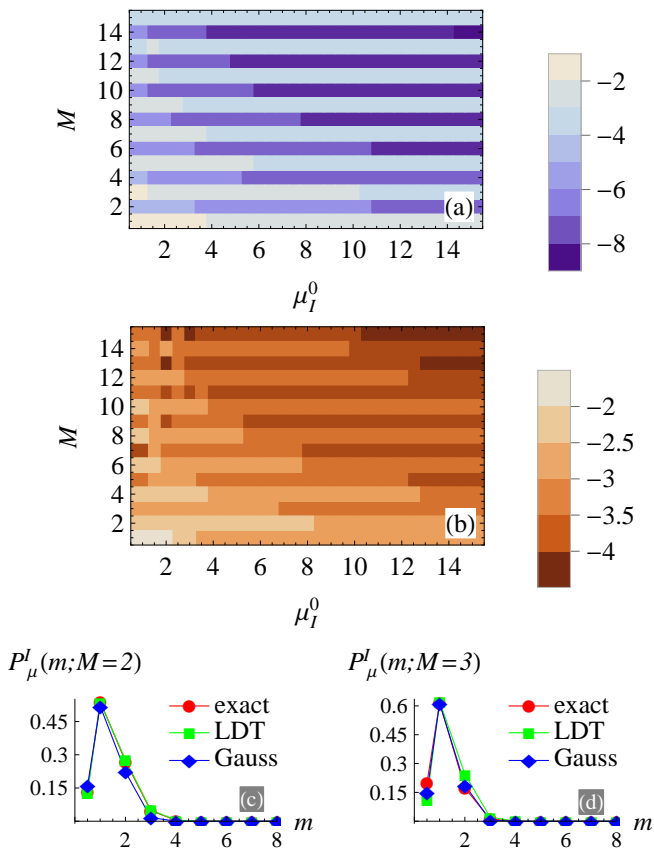}
\caption{(Color online) Distance between the exact distribution for $\mu_I(M)$ and those obtained by LDT approximation and the Gaussian approximation, and representative distributions for model I. Logarithm (base $e$) of the Jensen-Shannon divergence between (a) the exact distribution and the LDT approximation and (b) between the exact distribution and a Gaussian that has the same mean and variance as the exact distribution, calculated via Eq. (\ref{eqn:JSd}). (c-d) representative plots of the exact distribution as well as the LDT and Gaussian approximations for $\mu_{\mathrm{I}}^0=1$, $M=2$ (c), and $M=3$ (d).}
\label{fig:model1_exc_LDT}
\end{figure}

In Fig. \ref{fig:model1_exc_LDT}(a-b), we display the logarithm (base $e$) of the distance between $P_{\mu}^{\mathrm{I, exc}}$ and $P_{\mu}^{\mathrm{I, LDT}}$ (panel (a)), as well as the distance between $P_{\mu}^{\mathrm{I, exc}}$ and a Gaussian distribution with the same mean and variance as the exact distribution (panel (b)), using $\ln(\mathrm{JSd})$ as a measure (we integrate the Gaussian over $[L_m,R_m]$ and normalize, exactly in the same manner as the LDT probability density in Eq. (\ref{eqn:model1_pdf})). Darker shades correspond to lower values, as indicated in the color bar. We note that, on average, the LDT approximation is closer to the exact distribution than the Gaussian. We also note that the LDT approximation is much more accurate at even values of $M$ as opposed to odd values. This is the due to the fact that the sample average tends to be more biased when the sample set contains an odd number of elements than even. (An intuitive way to think about this is to consider coin tosses where heads count 1/2 and tails count -1/2, and we average over the results after $M$ consecutive tosses to produce an outcome. Note that when $M$ is odd, the outcome from a single round can never be zero). In Fig. \ref{fig:model1_exc_LDT}(c-d) we display example plots of the exact distribution of $\mu(M)$ as well as the LDT and Gaussian distributions for $\mu_{\mathrm{I}}^{0}=1$ and $M=2$ (c), and $M=3$ (d).

\subsubsection{Numerical results for all models}

For the three models under consideration, we calculated the probability distributions $P_{\mu}^{i}(\mu;M)= A \exp(-MI_i(\mu))$ predicted by the LDT, by numerically computing $I_i$, and integrating $\exp(-MI_i(\mu))$ over the intervals $[n\Delta \mu, (n+1)\Delta \mu )$, where $\Delta \mu$ is the bin size used to calculate probability distributions from simulation data, and $n=0,1,2,...,n_{\mathrm{max}}$. This ensures that we can directly compare analytical and simulation results. The normalization constant $A$ is calculated by considering a sufficiently large domain of integration such that the probability of observing a value between $[(n_{\mathrm{max}}+1)\Delta \mu,\infty)$ is negligible.
\begin{figure}[ht]
\centering
\includegraphics[width=\columnwidth]{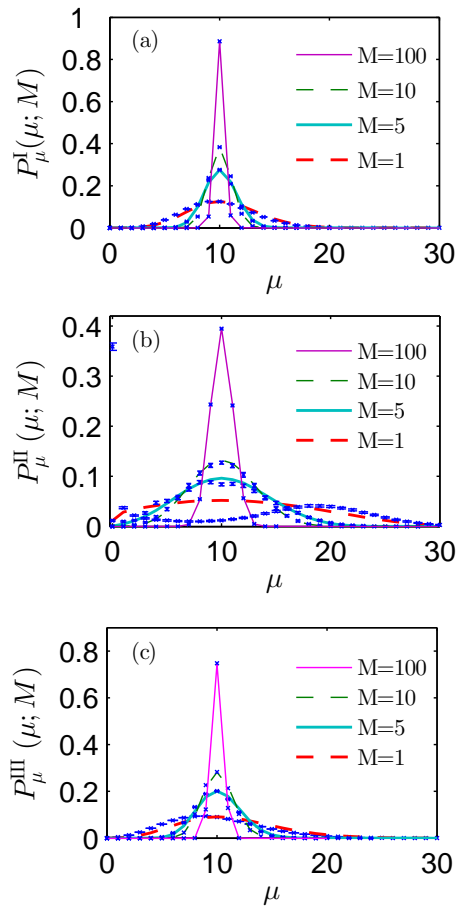}
\caption{(Color online) Comparison of simulation data with the predictions of LDT for the three models, for sample sizes $M = 1$ (thick-dashed, red), $M = 5$ (thick-solid, cyan), $M = 10$ (thin-dashed, green) and $M = 100$ (thin-solid, magenta). The simulation data is represented by crosses (blue) and error bars along y-axis, where most of the error bars are comparable to the size of data points. The agreement is good except for the smallest values of $M$, as expected. Panels show the comparison for model I (a), model II (b) and model III (c). Parameter values are (a): $g_1=10$, $d_1=1$; (b): $k_{\mathrm{on}}=0.1$, $k_{\mathrm{off}}=0.1$, $g_1=20$, $d_1=1$; (c): $g_1=10$, $d_1=10$, $g_2=10$, $d_2=1$.}
\label{fig:figure2}
\end{figure}

We performed stochastic simulations of the reaction systems illustrated in Fig \ref{fig:figure1}(a) by using the Gillespie algorithm~\cite{gillespie_exact_1977} and obtained the steady state probability distributions for the number of output molecules. In Fig. \ref{fig:figure2}(a-c), we demonstrate how the probability distributions obtained via the LDT converge to the simulation results as a function of the sample size $M$, with a set of parameter values $g_{i,j}$ and $d_{i,j}$ for which all models have the same mean number of output molecules $\mu_i^0=10$. For models I and III, we observe that the convergence is quick such that the LDT predictions and simulation results are hardly distinguishable for $M\geq 5$. For model II, we find that the convergence is less rapid and is attained for $M \geq 10$. For our choice of parameters, the simulation data for $M=1$ shows two peaks as the gene switches between the on and off states slowly compared to the rate of transcription, and the non-population averaged probability distribution retains maximal information about the process. The peak at $\mu=0$ cannot be seen in the LDT prediction (red line in Fig. \ref{fig:figure2}(b)). The poor agreement between the LDT predictions and simulation results for small $M$ partly lies in the fact that when $I_i$ can be calculated via the Legendre-Fenchel transform, it has to be a convex function such that $P_{\mu}^{i}(\mu;M)$ can never have two peaks~\cite{touchette_large_2009}.

We see that the distributions for different models are significantly different even though they have the same mean, and the distribution corresponding to $I_{\mathrm{II}}$ has larger variance. The probability distributions in Fig. \ref{fig:figure2}(a-c) are slightly asymmetric around the mean for small values of $M$, reflecting the skewness of the underlying distribution for a single cell. For large values of the mean expression level, we expect the asymmetry to disappear and the LDT predictions and simulation results to agree for even smaller values of $M$. When the mean number of molecules is sufficiently large, their probability distribution can be well approximated by a normal distribution, and the quadratic term in $I_i(\mu)$ alone can account for the fluctuations around the mean.

\subsubsection{Asymptotic behavior of $I_{\mathrm{II}}$ and $I_{\mathrm{III}}$}

For models II and III, the equations for $k^*$ are transcendental, which makes it difficult to find analytical solutions valid for all parameter values. Here we state our results on the behavior of $I_{\mathrm{II}}(\mu)$ and $I_{\mathrm{III}}(\mu)$ in the limit $\mu \to \infty$ where Eq. (\ref{eqn:for_the_root}) can be approximately solved for $k^*$. As we show in Appendix \ref{sec:appendix_1}, using approximate forms of Eqs. (\ref{eqn:equation_for_lambda}) and (\ref{eqn:for_the_root}), we obtain the following asymptotic results for $I_{\mathrm{II}}(\mu)$ and $I_{\mathrm{III}}(\mu)$
\begin{align}
I_{\mathrm{II}}(\mu) &\simeq \mu \ln \left( \frac{\mu}{\mu_{\mathrm{II}}^0\xi} \right) - \frac{\mu}{\xi}\left( \frac{k_{\rm on}+k_{\rm off}}{k_{\rm on}} \right) \nonumber \\ &+ \frac{k_{\rm off}}{d_1}\ln\left( \frac{\mu }{\mu_{\mathrm{II}}^0 \xi}\right) - \ln \xi^{\prime} \label{eqn:I_2_main}\\
&\simeq \mu \ln \mu , \\
I_{\mathrm{III}}(\mu) &\simeq \mu \ln \left( \ln \mu \right) - g_{1,2} \int_0^{\ln \mu} dv e^{v} \frac{\gamma(d_{1,2},v)}{v^{d_{1,2}}} \label{eqn:I_3_main} \\
&\simeq  \mu \ln \left( \ln \mu \right), \label{eqn:I_3_approximation_main}
\end{align}
where $\xi$ and $\xi^{\prime}$ are given by Eqs. (\ref{eqn:xi}) and (\ref{eqn:xi_prime}), respectively, and Eq. (\ref{eqn:I_3_approximation_main}) holds due to the inequality (see Appendix \ref{sec:appendix_1})
\begin{gather*}
\int_{0}^{\ln \mu} dv e^v \frac{d_{1,2}\gamma(d_{1,2},v)}{v^{d_{1,2}}} \leq \int_{0}^{\ln \mu} dv e^v  = \mu-1,
\end{gather*}
Note that when $k_{\rm off}=0$, we have $\xi=1$, $\ln \xi^{\prime}=-g_{1,1}=-\mu_{\mathrm{I}}^0$ and $I_{\mathrm{II}}(\mu)$ reduces to $I_{\mathrm{I}}(\mu)$ as expected.

\subsection{Remarks on several extensions of the basic models}
As we finish this section, we would like to note that the models considered above are the most basic models for gene expression, and it is instructive to discuss how the distribution of sample means would differ for several extensions of these basic models. As there has been considerable interest in obtaining analytical solutions for gene regulation networks, probability generating functions for a number of models are available, and can be used to perform the analysis we presented above. Here, we would like to discuss two recent results from the literature, with which a link between our findings can be established. Vandecan and Blossey~\cite{vandecan_self-regulatory_2013} derived exact results for the gene gate model, which pertains to a self-regulatory gene. These authors considered an extension of model II, where the gene product either activates or inhibits its synthesis by influencing the state of the gene. In both cases, a gene product $P$ emerges from the reaction $G_1\xrightarrow{g_1} G_1+P$, where $G_1$ is the first state of the gene and $g_1$ is the rate constant. $P$ decays over time with the rate constant $d_1$, independent of the state of the gene. The product $P$ then switches the state of the gene via the reaction $G_1+P\xrightarrow{k_{21}}G_2+P$. In the activator case, $G_2$ state decays along with the production of a $P$, that is, $G_2\xrightarrow{k_{12}} P$, resulting in the production of $P$ at an increased rate if $k_{12}$ is greater than $g_1$. In the repressor case, $G_2$ decays to $G_1$ via $G_2 \xrightarrow{k_{12}} G_1$, resulting in a decrease in the rate of production of $P$, as $G_2$ is analogous to the off state of the gene. The probability generating function for the activator case was found to be~\cite{vandecan_self-regulatory_2013}
\begin{gather}
Q_{\mathrm{A}}(z)= {}_1 F_{1}\left( a, b; c( z(k_{21}+d_1)-d_1)\right) / C_{\mathrm{A}}
\label{eqn:self_reg_A}
\end{gather}
where $a={k_{12}g_1}/({g_1 d_1 + k_{21} k_{12}})$, $b=({g_{1}k_{21} + k_{12}d_1})/(k_{21}+d_1)^2$, $c=(g_1 d_1 + k_{21} k_{12})/(d_1 (k_{21}+d_1)^2)$, and $C_{\mathrm{A}}$ is the normalization factor, ensuring that $Q_{\mathrm{A}}(1)=1$. The analogous expression for the repressing case was shown to be
\begin{gather}
Q_{\mathrm{R}}(z)= {}_1 F_{1}\left( a, b; c( z(k_{21}+d_1)-d_1)\right) / C_{\mathrm{R}}
\label{eqn:self_reg_R}
\end{gather}
where $a=k_{12}/d_1$, $b=(k_{21}(g_1+k_{12})+d_1 k_{12})/(k_{21}+d_1)^2$, $c=g_1/(k_{21}+d_1)^2$, and $C_{\mathrm{R}}$ is the normalization factor. Comparing these generating functions to that of model II (given in Eq. (\ref{eqn:pec_yca_pgf})), we realize that the feedback mechanism introduced in the gene-gating model preserves the form of the generating function, and rescales the variable $z$. Therefore, we expect that the qualitative features of the distribution of the sample means in this extended model would be similar to those of model II, while the physical meaning of the parameters is clearly different.

In another recent work, Pendar et al.~\cite{pendar_exact_2013} used an elegant approach based on the partitioning of Poisson processes, and derived exact results for extensions of model III, including one where the mRNA needs to be processed for a number of steps before is it ready to be translated. In biological terms, these post-transcriptional modifications include polyadenylation, splicing and translocation, which have often been excluded from models of stochastic gene expression. The steady state probability generating function for this extension that accounts for multistep mRNA processing was shown to be
\begin{align}
Q_{\mathrm{multi}}&(z) = \nonumber \\ &\lim_{N\to \infty} \exp \left( N\left[ {}_1 F_{1}\left( \frac{g_{\mathrm{eq}}}{d_{p}}, \frac{d_{0}}{d_{p}}; \frac{g_{p}}{d_{p}}(z-1) \right) -1 \right] \right),
\label{eqn:model3_pendar}
\end{align}
where $g_{0}$ and $d_{0}$ are the production and degradation rates of a premature mRNA molecule, which is further processed for $r$ number of steps before it results in a mature mRNA that is translated at rate $g_{p}$. As far as the steady state behavior is concerned, the effective rate at which the premature mRNA is converted into a mature transcript is given by
%\begin{gather*}
%g_{\mathrm{eq}} = g_{0}\prod_{i=1}^{r} \frac{g_{i}}{g_{i}+d_{i}}, 
%\end{gather*}
$g_{\mathrm{eq}} = g_{0}\prod_{i=1}^{r} {g_{i}}({g_{i}+d_{i}})^{-1}$
where $g_{i}$ is the rate at which the $i^{\mathrm{th}}$ intermediate mRNA is converted into the next form, and $d_{i}$ is the decay rate of each intermediate form. Upon taking the limit, Eq. (\ref{eqn:model3_pendar}) can also be expressed as 
\begin{align*}
Q_{\mathrm{multi}}(z) &= \exp \Bigg[ g_{\mathrm{eq},0} g_{p,p}(z-1) \\ &\times {}_2F_{2}\left( 1,1; 2,d_{0,p}+1; g_{p,p} (z-1)\right)  \Bigg],
\end{align*}
where ${}_2F_{2}$ is a generalized hypergeometric function~\cite{gradshtein_table_2007}, and we used our shorthand notation $g_{i,j}=g_{i}/d_{j}$. In Appendix \ref{sec:appendix_2}, we show that the generating function of model III can also be put in this form, with $g_{\mathrm{eq}}$ replaced by $g_{1}$ and $g_{p}$ replaced by $g_{2}$, along with the corresponding decay rate constants. Therefore, the qualitative behavior of the sample mean from this extended model would be the same as what we have found for model III, with a reinterpretation of parameters. Lastly, we would like to make a remark that the time dependent expressions for the generating functions for these extended models were also provided in refs.~\cite{pendar_exact_2013} and \cite{vandecan_self-regulatory_2013}.

\section{Discussion and conclusions}
\label{sec:discussion}
Quantifying the fluctuations in the expression levels of mRNA and proteins is a contemporary challenge in cell biology. It is commonly observed that a colony of genetically identical cells is quite heterogeneous in its gene expression profile. However, finding the reason for this diversity and whether it has any biological function requires a significant amount of data collection and analysis. Observing mRNA and protein levels at the single cell level, which in principle gives us maximal information about the gene expression mechanism, is becoming increasingly common practice. Nevertheless, conclusive analyses demand large volumes of data that often requires a disproportionately large amount of effort and time to obtain. In this work, we investigated the extent to which population averaged measurements, as opposed to single cell level measurements, can reflect the underlying gene expression mechanism.

To explore our idea, we considered three simple, but fundamental gene expression networks that predict the number distribution of mRNA and proteins for constitutively, as well as transiently expressed genes. Constitutive expression models are often relevant for gene expression in prokaryotes, whereas transient expression models have an extra degree of freedom to account for gene switching due to chromatin remodeling in eukaryotic cells. Using exact solutions of these models and the Large Deviation Theory (LDT), we calculated the distribution of gene expression products averaged over a population of size $M$, and its first two moments.

To investigate the effect of the sample size, $M$, and the finite number of independent measurements of the sample mean, $N$, we calculated the uncertainty in the first two moments as a function of $M$ and $N$. Using the first two moments, we also derived an expression for the relative uncertainty in the Fano factor, a commonly used quantity to characterize burstiness of gene expression (see Eq. (\ref{eqn:fano_uncertainty})). We separated the model dependent and independent parts of this expression, and provided specific results for models II and III (see Fig. \ref{fig:var_var}).     

Since the LDT only provides an approximation to the distribution of the population average, we compared its predictions with exact results and stochastic simulations to determine how large the population should be, before one can apply the theory. For the simplest model considered, model I, we quantified the difference between the LDT prediction and exact distributions using the Jensen-Shannon divergence, finding good agreement between the two (see Fig. \ref{fig:model1_exc_LDT}). For all models, we compared the predictions of the LDT with simulation results for small sample sizes where the distribution of the population average is non-Gaussian, and found reasonable agreement as $M$ increases (see Fig. \ref{fig:figure2}). 

In addition, by performing an asymptotic analysis we showed that the behavior of the rate functions $I_i(\mu)$ depend on the number of stages in which the gene product is synthesized. For models I and II, which are used to model the synthesis of mRNA in a single step, the rate function has the asymptotic form $\simeq \mu \ln \mu$, whereas for model III, which describes the synthesis of a protein in a two step reaction, the rate function goes as $\simeq \mu \ln(\ln \mu)$. This suggests that there is a relation between the number of steps in which the final product is synthesized, and the asymptotic form of the rate function, in the form of nested logarithms. In our future work, we would like to further explore this by calculating the rate functions for a gene expression cascade consisting of $n$ steps.

We hope that our theoretical results can be useful in estimating the approximate number of cells for which the distribution of sample means is still informative. In cases where measurement precision allows one to extract information from the variance of the sample mean, characterized by Eqs. (\ref{eqn:M_max}) and (\ref{eqn:fano_uncertainty}), using larger values of $M$ can reduce the amount of data that needs to be stored and processed in order to extract parameters such as transcription and translation rates via model fitting. 

Lastly, we discussed how the analysis results we presented for the three basic models would differ if extensions of these models are considered. To this end, we considered the recent results derived by Vandecan and Blossey~\cite{vandecan_self-regulatory_2013} for a self-regulating gene, and by Pendar et al.~\cite{pendar_exact_2013} for cases where mRNA needs to be processed in multiple steps before resulting in a transcript that can be translated. We pointed out that the steady state generating functions of these extensions are essentially the same as those for models II and III, making it possible to argue how predictions for the sample mean would differ for more complex cases, via a reinterpretation of parameters in simpler models. It would be highly desirable to include in our analysis an extension of model II, which accounts for the translation of mRNA such that the distribution of protein numbers can be considered. Despite the obvious biological significance of this extension, to the best of our knowledge, exact results for the generating function are not available in the literature. Nevertheless, in previous work by Assaf et al.~\cite{assaf_determining_2011} and Newby~\cite{newby_isolating_2012}, the authors derived expressions for the probability distribution for an extension of model II with two stages and nonlinear feedback, using a WKB approximation. Therefore, results obtained in those studies can provide a base for approximating the distribution of population averaged mRNA and protein levels via the method outlined in this work.

\begin{acknowledgments}
We would like to thank three anonymous reviewers for their constructive criticism that improved our original manuscript. This research was supported in part by the World Premier International Research Center (WPI) Initiative of the Ministry of Education, Culture, Sports, Science and Technology (MEXT) of Japan.
\end{acknowledgments}

\appendix

\section{Moments of a sample mean}
\label{sec:appendix_0}
In order to calculate moments of $\mu(M)$, we first calculate the moments of $Y(M)=M \mu(M)=\sum_i X_i$, a random variable obtained by summing up $M$ independently and identically distributed random variables. Moments of $Y(M)$ can be expressed in terms of the moments of the individual random variables $X_i$. Using the formula obtained in Packwood~\cite{packwood_moments_2011}, we deduce that the first four moments of $Y$ are related to those of $X$ through
\begin{align}
\left\langle {Y(M)}^2 \right\rangle &= M\left\langle {X}^2 \right\rangle + M(M-1){\left\langle {X} \right\rangle}^2, \\
\left\langle {Y(M)}^3 \right\rangle &= M \left\langle {X}^3 \right\rangle + 3M(M-1)\left\langle {X}^2 \right\rangle  \left\langle {X} \right\rangle \nonumber \\ &+ M(M-1)(M-2) {\left\langle {X} \right\rangle}^3, \\
\left\langle {Y(M)}^4 \right\rangle &= M \left\langle {X}^4 \right\rangle + 4M(M-1) \left\langle {X}^3 \right\rangle \left\langle {X} \right\rangle \nonumber \\ &+ 3M(M-1) {\left\langle {X}^2 \right\rangle}^2 \nonumber \\ &+ 6M(M-1)(M-2) \left\langle {X}^2 \right\rangle {\left\langle {X} \right\rangle}^2 \nonumber \\ &+ M(M-1)(M-2)(M-3) {\left\langle {X} \right\rangle}^4.
\end{align}
Using these relations, one can immediately calculate the moments of $\mu(M)$ via
\begin{gather}
\left\langle {\mu(M)}^{\ell} \right\rangle = \frac{1}{M^{\ell}} \left\langle {Y(M)}^{\ell} \right\rangle.
\end{gather}

\section{Derivation of the asymptotic forms for $I_{\mathrm{II}}(\mu)$, $I_{\mathrm{III}}(\mu)$}
\label{sec:appendix_1}
Here we provide the details of the calculations that lead to the asymptotic expressions given in Eqs. (\ref{eqn:I_2_main}) and (\ref{eqn:I_3_main}).

For model II, substituting the pgf given in Eq. (\ref{eqn:pec_yca_pgf}) in Eq. (\ref{eqn:for_the_root}), we obtain the following equation that needs to be solved to obtain $k^*$
\begin{gather}
\frac{e^k {}_1F_1\left(\frac{k_{\textrm{on}}}{d_1}+1, \frac{k_{\textrm{on}}+k_{\textrm{off}}}{d_1}+1, g_{1,1}(e^k-1) \right)}{{}_1F_1\left(\frac{k_{\textrm{on}}}{d_1}, \frac{k_{\textrm{on}}+k_{\textrm{off}}}{d_1}, g_{1,1}(e^k-1) \right)} = \frac{\mu}{\mu_{\mathrm{II}}^0},
\label{eqn:root_for_II}
\end{gather}
where 
\begin{gather}
\mu_{\mathrm{II}}^0=\frac{g_{1,1} k_{\rm on}}{(k_{\rm on} + k_{\rm off})}.
\end{gather}

Next, we examine the behavior of $I_{\mathrm{II}}(\mu)$ as $\mu \to \infty$, while $\mu_{\mathrm{II}}^0$ is finite. We observe that the left hand side of Eq. (\ref{eqn:root_for_II}) is monotonically increasing in $k$. Therefore, if we seek a solution as $\mu \to \infty$, we can as well consider replacing the left hand side of Eq. (\ref{eqn:root_for_II}) by its leading term as $k \to \infty$, that is
\begin{gather}
\frac{e^k {}_1F_1\left(\frac{k_{\textrm{on}}}{d_1}+1, \frac{k_{\textrm{on}}+k_{\textrm{off}}}{d_1}+1, g_{1,1}(e^k-1) \right)}{{}_1F_1\left(\frac{k_{\textrm{on}}}{d_1}, \frac{k_{\textrm{on}}+k_{\textrm{off}}}{d_1}, g_{1,1}(e^k-1) \right)} \simeq \xi e^k,
\label{eqn:root_for_II_approx}
\end{gather}
where
\begin{gather}
\xi =\frac{ \Gamma\left( 1+\frac{k_{\rm on}+k_{\rm off}}{d_1} \right) \Gamma\left(\frac{k_{\rm on}}{d_1} \right)}{ \Gamma\left( 1+\frac{k_{\rm on}}{d_1} \right) \Gamma\left(\frac{k_{\rm on}+k_{\rm off}}{d_1} \right)}.
\label{eqn:xi}
\end{gather}
Hence, we have
\begin{gather}
k^* \simeq \ln \frac{\mu}{\mu_{\mathrm{II}}^0 \xi}.
\end{gather}
In the same manner, finding the leading term of $\lambda_2(k)$ as $k\to \infty$, we write
\begin{align}
\lambda_2(k) &\simeq \ln \left( \exp\left( g_{1,1}e^k \right) e^{-\frac{k_{\rm off}}{d_1}k}\left[ \xi^{\prime} + O(e^{-k}) \right]  \right) \nonumber \\
&\simeq g_{1,1}e^k - \frac{k_{\rm off}}{d_1}k + \ln \xi^{\prime},
\label{eqn:lambda_II_approx}
\end{align}
where 
\begin{gather}
\xi^{\prime} = \frac{e^{-g_{1,1}}}{ g_{1,1} ^{{k_{\rm off}}/{d_1}} } \frac{ \Gamma\left( \frac{k_{\rm on}+k_{\rm off}}{d_1} \right)}{\Gamma\left( \frac{k_{\rm on}}{d_1} \right)}.
\label{eqn:xi_prime}
\end{gather}
Substituting $k = k^*$ in Eq. (\ref{eqn:lambda_II_approx}), we deduce
\begin{align}
I_{\mathrm{II}}(\mu) &\simeq \mu \ln \left( \frac{\mu}{\mu_{\mathrm{II}}^0\xi} \right) - \frac{\mu}{\xi}\left( \frac{k_{\rm on}+k_{\rm off}}{k_{\rm on}} \right) \nonumber \\ &+ \frac{k_{\rm off}}{d_1}\ln\left( \frac{\mu }{\mu_{\mathrm{II}}^0 \xi}\right) - \ln \xi^{\prime}.
\end{align}
Note that when $k_{\rm off}=0$, we have $\xi=1$, $\ln \xi^{\prime}=-g_{1,1}=-\mu_{\rm I}^0$ and $I_{\mathrm{II}}(\mu)$ reduces to $I_{\mathrm{I}}(\mu)$ as expected. This is because the approximations in Eqs. (\ref{eqn:root_for_II_approx}) and (\ref{eqn:lambda_II_approx}) become exact when $k_{\rm off}=0$. 

Next, we consider model III. In this case Eq. (\ref{eqn:for_the_root}) becomes 
\begin{gather}
g_{2,1}^{-1}\left(q+g_{2,2}\right)e^{q} \frac{ \gamma\left( d_{1,2}, q \right) }{ q^{d_{1,2}}} = \frac{\mu}{\mu_{\mathrm{III}}^0},
\label{eqn:for_the_root_N_2}
\end{gather}
where $q=g_{2,2}(e^k-1)$ and $\mu_{\mathrm{III}}^0=g_{1,1}g_{2,2}$.

In approximating $I_{\mathrm{III}}(\mu)$ at large values of $\mu$, we note that the function in the left hand side of Eq. (\ref{eqn:for_the_root_N_2}) is monotonically increasing with respect to $q$, similar to what we observed for model II. Therefore, large values of $\mu$ implies large values of $q$. Noting that $\lim_{z \to \infty} \gamma(a,z) = \Gamma(a)$, in the limit $q\to\infty$ we approximate the incomplete gamma function by a constant and write
\begin{gather}
g_{2,1}^{-1}\left(q+g_{2,2}\right) e^q \frac{\Gamma(d_{1,2})}{q^{d_{1,2}}} \simeq \frac{\mu}{\mu_{\mathrm{III}}^0},
\label{eqn:for_the_root_N_2_2}
\end{gather}
Taking the logarithm of both sides of Eq. (\ref{eqn:for_the_root_N_2_2}), we obtain
\begin{gather}
\ln \Gamma(d_{1,2}) - d_{1,2}\ln q + \ln \left(\frac{q}{g_{2,1}}+d_{1,2}\right)+ q \nonumber \\ \simeq \ln \left( g_{2,1}\frac{\mu}{\mu_{\mathrm{III}}^0} \right).
\end{gather}
As $q\to \infty$, the terms that are logarithmic in $q$ will be negligible compared with the linear term, allowing us to further approximate the above formula, and write
\begin{gather}
q \simeq (g_{2,1}\frac{\mu}{\mu_{\mathrm{III}}^0}).
\label{eqn:for_the_root_N_2_3}
\end{gather}
Therefore, as $\mu \to \infty$, the root $k^*$ can be approximated by
\begin{align}
k^* &\simeq  \ln \left( 1 + g_{2,2}^{-1}\ln\left(g_{2,1} \frac{\mu}{\mu_{\mathrm{III}}^0} \right) \right), \nonumber \\
&\simeq \ln \left( \ln \mu \right).
\end{align}
Substituting $k^*$ in Eq. (\ref{eqn:equation_for_I}) we obtain the function $I_{\mathrm{III}}(\mu)$ as
\begin{gather}
I_{\mathrm{III}}(\mu) \simeq \mu \ln \left( \ln \mu \right) - g_{1,2} \int_0^{\ln \mu} dv e^{v} \frac{\gamma(d_{1,2},v)}{v^{d_{1,2}}} .
\label{eqn:I_N_2_1}
\end{gather}
We note that the integral in Eq. (\ref{eqn:I_N_2_1}) can be put in the following form
\begin{gather}
g_{1,1} \int_{0}^{\ln \mu} dv e^v \frac{d_{1,2}\gamma(d_{1,2},v)}{v^{d_{1,2}}},
\end{gather}
where 
\begin{gather}
\frac{d_{1,2}\gamma(d_{1,2},v)}{v^{d_{1,2}}} \leq 1,
\end{gather}
for all values of $v$ and $d_{1,2}$. Therefore, we have
\begin{gather}
\int_{0}^{\ln \mu} dv e^v \frac{d_{1,2}\gamma(d_{1,2},v)}{v^{d_{1,2}}} \leq \int_{0}^{\ln \mu} dv e^v  = \mu-1,
\end{gather}
which means that the leading term in Eq. (\ref{eqn:I_N_2_1}) is $\mu \ln(\ln \mu)$ as $\mu\to \infty$, such that we have
\begin{gather}
I_{\mathrm{III}}(\mu) \simeq \mu \ln(\ln \mu).
\end{gather}

\section{Probability generating function for model III}
\label{sec:appendix_2}

Here we outline the derivation of Eq. (\ref{eqn:CME_model_III}), which is a special case of the joint probability generating function for model II. We follow the approach taken previously by Shahrezaei and Swain~\cite{shahrezaei_analytical_2008}; however, we present a more general result without making the approximation that the mRNA lifetime is negligible compared to that of the protein, and obtain the joint probability generating function as well as the marginal generating functions. For a comprehensive mathematical analysis of this model we refer the reader to the recent work by Bokes et al.~\cite{bokes_exact_2012}.

Using Eqs. (\ref{eqn:pgf}) and (\ref{eqn:CME_model_III}), it is deduced that the steady state probability generating function obeys the following partial differential equation
\begin{gather}
-d_{1,2} \left( g_{2,1}(1+u_1)u_2 - u_1 \right)\frac{\partial Q}{\partial u_1} +  u_2 \frac{\partial Q}{\partial u_2} = g_{1,2} u_1 Q,
\label{eqn:generating_N_2}
\end{gather}
where $u_i=z_i-1$. As Eq. (\ref{eqn:generating_N_2}) is a first-order linear partial differential equation, method of characteristics can be applied to solve it~\cite{evans_partial_2010}. Letting $s$ be the parameter of the characteristic curves, we have the following set of ordinary differential equations
\begin{align}
\frac{d u_1}{ds} &= -d_{1,2} \left( g_{2,1}(1+u_1)u_2 - u_1 \right), \label{eqn:char_u1}\\
\frac{d u_2}{ds} &= u_2, \\
\frac{d Q}{ds} &= g_{1,2} u_1 Q. \label{eqn:char_Q} 
\end{align}
The characteristic curve for $u_2$ is found by direct integration
\begin{gather}
u_2(s) = u_2(s_0)e^s.
\label{eqn:u_2}
\end{gather}
Substituting $u_2$ in Eq. (\ref{eqn:char_u1}) and using the integrating factor method, we obtain the characteristic curve for $u_1$ as
\begin{align}
u_1(s) &= e^{d_{1,2}(s-g_{2,1} u_2(s_0)e^s)} \Bigg[ u_1(s_0) \nonumber \\ &- \sum_{n=0}^{\infty} \frac{\left( g_{2,2} u_2(s_0) \right)^{n+1}}{n!(n-d_{1,2}+1)} e^{s(n-d_{1,2}+1)} \Bigg].
\label{eqn:u_1}
\end{align}
Note that the constant $u_1(s_0)$ will contain a diverging component if $d_{1,2}$ is an integer and cancel out the diverging term of the series. Finally solving Eq. (\ref{eqn:char_Q}), we obtain
\begin{align}
&Q(s) = Q_{s_0} \nonumber \\& \times \exp\Big[ g_{1,2}\Big\{ u_1(s_0) \sum_{m=0}^{\infty} \frac{(-g_{2,2}u_2(s_0))^m}{m!(m+d_{1,2})} e^{s(m+d_{1,2})} \nonumber \\ &- \sum_{n=0}^{\infty} \frac{\left( g_{2,2} u_2(s_0) \right)^{n+1}}{n!(n-d_{1,2}+1)}  \sum_{m=0}^{\infty} \frac{(-g_{2,2}u_2(s_0))^m}{m!(m+n+1)} e^{s(m+n+1)}   \Big\} \Big],
\label{eqn:Q_first}
\end{align}
where $Q_{s_0}$ is a constant with respect to $s$, and we made use of the identity
\begin{gather}
\int_{s_0}^s dk e^{ak-be^k} = \sum_{m=0}^{\infty} \frac{(-b)^m}{m!(m+a)}\left( e^{s(m+a)}-e^{s_0(m+a)} \right),
\end{gather}
which is straightforward to prove by expanding the integrand in Taylor series. We proceed by performing algebraic operations on Eq. (\ref{eqn:Q_first}). Lumping all the terms that are constant with respect to $s$ in the undetermined constant $Q_{s_0}$, and finally expressing $u_1(s_0)$ and $u_2(s_0)$ in terms of $u_1$, $u_2$ and $s$ (via Eqs. (\ref{eqn:u_2}-\ref{eqn:u_1})) and substituting them back in, we obtain
\begin{align}
Q(&u_1,u_2) = Q_{s_0} \exp \Bigg[ g_{1,2}\Bigg\{ u_1 e^{g_{2,2}u_2}\sum_{n=0}^{\infty} \frac{(-g_{2,2} u_2)^n}{n!(n+d_{1,2})} \nonumber \\ &+ \sum_{n=0}^{\infty} \frac{( g_{2,2}u_2)^{n+1}}{n! } \sum_{m=0}^{\infty}\frac{(- g_{2,2}u_2)^m}{m! (m+d_{1,2}) (m+n+1)} \Bigg\} \Bigg].
\label{eqn:Q_second}
\end{align}
%Using the definitions of the upper and lower incomplete gamma functions~\cite{abramowitz_handbook_1964},
%\begin{align*}
%\Gamma(a,z)&=\int_z^{\infty} dt e^{-t} t^{a-1}, & \Gamma(a,0) &= \Gamma(a), \\
%\gamma(a,z)&=\int_0^z dt e^{-t} t^{a-1},  & \Gamma(a,z) + \gamma(a,z) &= \Gamma(a),
%\end{align*}
In a recent work, Bokes et al.~\cite{bokes_exact_2012} also obtained this exact result using a different approach (see Eqs. (22) and (26) of ref.~\cite{bokes_exact_2012}). Eq. (\ref{eqn:Q_second}) can also be expressed in the following form
%\begin{align}
%Q(u_1,u_2) &= Q_{s_0} \exp \Bigg[ g_{1,2}\Bigg\{ u_1 e^{ g_{2,2}u_2}  \frac{\gamma(d_{1,2},g_{2,2} u_2)}{( g_{2,2} u_2)^{d_{1,2}}}  \nonumber \\ &+ \sum_{n=0}^{\infty} \frac{( g_{2,2}u_2)^{n+1}}{n! (n-d_{1,2}+1)} \Big( \frac{\gamma(d_{1,2},g_{2,2} u_2)}{( g_{2,2} u_2)^{d_{1,2}}} \nonumber \\ &- \frac{\gamma(n+1, g_{2,2} u_2)}{( g_{2,2} u_2)^{n+1}} \Big) \Bigg\} \Bigg], \\
%&= Q_{s_0} \exp \Bigg[ g_{1,2}\Bigg\{ u_1 e^{ g_{2,2}u_2}  \frac{\gamma(d_{1,2},g_{2,2} u_2)}{( g_{2,2} u_2)^{d_{1,2}}} \nonumber \\ &+ \int_0^{g_{2,2}u_2} dv e^{v} \frac{\gamma(d_{1,2},v)}{v^{d_{1,2}}} \Bigg\} \Bigg],
%\label{eqn:Q_third}
%\end{align}
\begin{align}
Q(u_1,u_2)&= \nonumber \\ &Q_{s_0} \exp \Bigg[ g_{1,2}\Bigg\{ u_1 \frac{e^{ g_{2,2}u_2}}{g_{2,2}^{d_{1,2}}}\int_0^{g_{2,2}u_2}e^{-y}y^{d_{1,2}-1}dy\nonumber \\&
 +\int_0^{g_{2,2}u_2} \frac{e^vdv}{v^{d_{1,2}}}\int_{0}^{v} e^{-t}t^{d_{1,2}-1}dt \Bigg\} \Bigg]
\label{eqn:Q_third}
\end{align}
As $P$ is a normalized probability distribution, $Q(0,0)$ must be equal to 1. Setting $u_1,u_2=0$, we deduce $Q_{s_0}=1$. Furthermore, we expect the marginal probability distribution $P(n_1)$ to be a Poisson distribution as $n_1$ is not consumed when $n_2$ is synthesized. Noting that $Q(u_1)=Q(u_1,0)$, we evaluate Eq. (\ref{eqn:Q_second}) at $u_2=0$ and find
\begin{gather}
Q(u_1) = e^{g_{1,1} u_1}, 
\end{gather}
which is indeed the generating function of the Poisson distribution with parameter $g_{1,1}$ (see Eq. (\ref{eqn:poisson_pgf})). Setting $u_1=0$, we obtain the generating function for the probability distribution of $n_2$ as
\begin{align}
Q(u_2) = \exp \Bigg[ g_{1,2} \int_0^{g_{2,2}u_2} dv e^{v} \frac{\gamma(d_{1,2},v)}{v^{d_{1,2}}} \Bigg].
\label{eqn:Q_marginal_u2_2}
\end{align}
This function can also be written in the following form 
\begin{align}
Q(u_2) = \exp \Bigg[ \mu_{\mathrm{III}}^{0}u_2 \ {}_2F_{2}\left( 1,1; 2,d_{1,2}+1; g_{2,2}u_2 \right)  \Bigg],
\label{eqn:Q_marginal_u2_2_hyper}
\end{align}
by expressing the double sum in the right hand side of Eq. (\ref{eqn:Q_second}) in terms of the hypergeometric function ${}_2F_{2}(a_1,a_2;b_1,b_2;z)$. A set of useful identities involving hypergeometric functions can be found in chp. 9 of ref.~\cite{gradshtein_table_2007}. Note that this expression is identical to the result obtained by Pendar et al.~\cite{pendar_exact_2013}, who developed a new approach using the partitioning of poisson processes (see Eq. (6) of ref.~\cite{pendar_exact_2013}, after performing the limit).

\bibliography{BB_ZK_biblio}

\end{document}